\documentclass[10pt,journal,doublecolumn, a4paper]{IEEEtran}

\IEEEoverridecommandlockouts

\usepackage{graphics} 
\usepackage{epsfig} 
\usepackage[numbers,sort&compress]{natbib}
\usepackage{multirow}
\usepackage{booktabs,graphicx}
\usepackage{rotating}
\usepackage[left=0.58in, right=0.58in, top=2 cm]{geometry} 
\usepackage{amsmath} 
\usepackage{amssymb}  
\usepackage{colortbl}
\usepackage{threeparttable}
\newcommand{\figwidth}{0.9\columnwidth}
\newcommand{\comment}[1]{}
\usepackage{flushend}
\usepackage{xcolor}

\usepackage{caption}

\usepackage[boxed,ruled,vlined,linesnumbered,commentsnumbered, noresetcount]{algorithm2e}
\usepackage{algorithmic}
\usepackage{bm}

\allowdisplaybreaks

\renewcommand{\baselinestretch}{.92}

\begin{document}

\title{Concatenated Code Design for Constrained DNA Data Storage with Asymmetric Errors}

\author{Yixin Wang,~
Li Deng,~
Md. Noor-A-Rahim,~
Erry Gunawan,~
Yong L. Guan,~
Zhi P. Shi,~
Chueh L. Poh
\thanks{Yixin Wang, Erry Gunawan  and Yong Liang Guan are with the School of Electrical and Electronic Engineering (EEE), Nanyang Technological University (NTU), Singapore (E-mail: {\tt \{ywang065, egunawan, eylguan\}@ntu.edu.sg}). Li Deng and Zhi Ping Shi are with the National Key Laboratory of Science and Technology on Communications, University of Electronic Science and Technology of China (UESTC) (E-mail: {\tt dengli@std.uestc.edu.cn, szp@uestc.edu.cn}) and  Li Deng is also with the School of Electrical and Electronic Engineering (EEE), Nanyang Technological University (NTU), Singapore.  Md. Noor-A-Rahim is  with the School of Computer Science and IT, University College Cork, Ireland (E-mail: {\tt m.rahim@cs.ucc.ie}). Chueh Loo Poh is  with the  Department of Biomedical Engineering, National University of Singapore (NUS), Singapore (E-mail: {\tt poh.chuehloo@nus.edu.sg}). (Corresponding author: Chueh Loo Poh)
}
}

\maketitle


\begin{abstract}
 DNA Data storage has recently attracted much attention due to its durable preservation and extremely high information density (bits per gram) properties. In this work, we propose a hybrid coding strategy comprising of generalized constrained codes to tackle homopolymer (run-length) limit and a protograph based low-density parity-check (LDPC) code to correct asymmetric nucleotide level (i.e., A/T/C/G) substitution errors that may occur in the process of DNA sequencing. Two sequencing techniques namely, Nanopore sequencer and Illumina sequencer with their equivalent channel models and capacities are analyzed. A coding architecture is proposed to potentially eliminate the catastrophic errors caused by the error-propagation in the constrained decoding while enabling high coding potential. We also show the log likelihood ratio (LLR) calculation method for the belief propagation decoding with this coding architecture. The simulation results and the theoretical analysis show that the proposed coding scheme exhibits good bit-error rate (BER) performance and high coding potential ($\sim1.98$ bits per nucleotide).
\end{abstract}


\section{Introduction}

 



The current digital data storage systems, such as optical or magnetic data storage, are reaching a limit to store ever-growing data due to the limitation in density. In recent years, DNA has brought a new opportunity for building reliable long-term storage with high capacity. In a DNA-based data storage, two techniques, named DNA synthesis and DNA sequencing, are used to write data and read data, respectively. Before the synthesis of DNA strands, the binary data need to be mapped into the DNA units (e.g., nucleotides) such that a balanced GC content\footnote{The number of alphabet 'G' and 'C' against the length of a DNA sequence.} and homopolymer runs\footnote{The length of consecutively repetitive alphabets, also denoted as run-length.} less than 4nt can be achieved at the strand level. These characteristics are desired because synthesized DNA strands that do not satisfy the above mentioned biochemical constraints (i.e., DNA strands with high/low GC content or long homopolymer runs) are prone to sequencing errors \cite{ross2013characterizing}. Moreover, to ensure the data integrity, error correction codes are often incorporated to correct the errors caused by mutations. The existing error correction schemes for DNA data storage are mostly implemented at the strand-level, relying on an outer code to recover the erroneous strands based on other error-free strands (e.g., \cite{Erlich2017}). In other words, the existing works do not attempt to correct the erroneous strands even when there is a single nucleotide error. As such, this might cause high computation and time complexity in decoding of the outer codes. In the past experimental results \cite{Organick2018}, it is observed that one of the most significant error patterns in DNA data storage is the substitutions among nucleotides induced from the sequencing process. In addition, the substitution possibilities of four nucleotides are different for two widely utilized sequencers (i.e., Nanopore sequencers \cite{gabrys2017asymmetric} and Illumina’s NextSeq sequencers \cite{Organick2018}). Therefore, we are inspired to devise an error control strategy at the nucleotide-level to address such asymmetric errors, potentially assisting the outer decoder to recover the stored data. 

In this work, we design a hybrid coding scheme for DNA-based data storage to tackle asymmetric mutations while satisfying the biochemical constraints. We have developed a variable-length run-length limited (VL-RLL) code with its modified version for DNA mapping that results in a homopolymer-constrained DNA data storage system with high efficiency. In addition, we derive the channel capacities of the asymmetric Nanopore sequencer and Illumina sequencer. To tackle the asymmetric errors, we use a protograph low-density parity-check (LDPC) coding scheme at the near-nucleotide-level, correcting the channel errors before they expanding to more catastrophic errors in the constrained decoding. Besides, the Log-Likelihood Ratio (LLR) calculation method is described for the belief propagation decoding in the proposed coding scheme. 






\section{Preliminaries} 
\label{sec:Preliminary}

\subsection{VL-RLL codes for constrained DNA data storage}
\begin{figure}[htbp]
  \centering
  \includegraphics[width=\figwidth, height = 2.8cm  ]{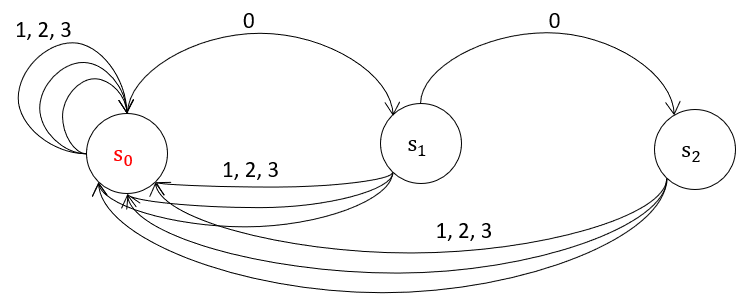}
  \caption{Finite state transition diagram of (4, 0, 2) constrained DNA storage.}
  \label{fig:FSTD}
\end{figure}

DNA data storage can be considered as a run-length limited (RLL) constrained system due to the maximum homopolymer run limit on DNA sequences. With the maximum homopolymer run to 3nt, DNA data storage becomes a $(4, 0, 2)$ constrained system, representing by a finite state transition diagram (FSTD) as Fig. \ref{fig:FSTD} \cite{8695057}. In this work, we employed the variable-length RLL codes rather than the fixed-length RLL codes since the former exhibits much lower coding complexity in order to achieve an equivalent coding rate with the latter. However, we found that the error-propagation in VL-RLL codes is severer than the fixed-length codes as the variable-length characteristics of the source words and codewords might bring in additional synchronization problems. To address this, we have devised a new effective coding scheme in Section. \ref{sec:Heterogeneous Coding Scheme}.

 Following the same approach as \cite{8695057}, we build a finite set ($\{1, 2, 3, 01, 02, 03, 001, 002, 003\}$) of which each element can be arbitrarily concatenated into long words that ultimately comply with the run-length limit. The concatenations of elements in this basic set can be used as the $(4, 0, 2)$ transition word set to establish the bijection with the source word set. For simplicity, we use the basic set as the transition word set mapping to the source data. With the assumption of i.i.d binary source data, the variable-length source words is assigned to the variable-length codewords via the Huffman approach \cite{huffman1952method}, attaining a bijective mapping as shown in Table. \ref{tab:VL-RLL}. The average coding potential (bits per symbol) thus is optimized to $1.976$,  calculated by \cite{steadman2013variable},

 
 

\begin{equation}  \label{eq:code rate}  
R=\frac{\sum_i{2^{-l_i}l_i}}{\sum_i{2^{-l_i}o_i}}
\end{equation}
\comment{
An example of encoding binary sequence to VL-RLL DNA sequence is shown as follows. Supposing we have a binary sequence $'1110$-$10$-$01$-$1101$-$00$-$111100'$. According to the mapping rule in Table. \ref{tab:VL-RLL}, we obtain the transition sequence $'03$-$3$-$2$-$02$-$1$-$001'$ . After differential precoding, the transition sequence is transited to the RLL sequence $'03$-$2$-$0$-$02$-$3$-$330'$. By mapping '0'$\to$'A', '1'$\to$'T', '2'$\to$'G', and '3'$\to$'C', we construct the final RLL DNA sequences 'ACGAAGCCCA', avoiding homopolymer longer than 3nt.
}

\begin{table}
\setlength{\extrarowheight}{1.5mm} 
\setlength{\tabcolsep}{0.9mm}
\centering 
\caption{VL-RLL Mapping Rule}\label{tab:VL-RLL}
\begin{threeparttable}
\begin{tabular}{|c|c|c|c|c|c|c|c|c|c|}
\hline
\textbf{Source word} &  $00$ & $01$ &  $10$ & $1100$ &  $1101$ & $1110$ &  $111100$ & $111101$ & $11111$\\ 
\hline
\textbf{Transition word} &  $1$ & $2$ &  $3$ & $01$ &  $02$ & $03$ &  $001$ & $002$ & $003$\\
\hline
\end{tabular}
\end{threeparttable}
\end{table}


\subsection{LDPC codes for DNA data storage}

\begin{figure}[htbp]
  \centering
  \includegraphics[width=\figwidth, height = 4.5cm]{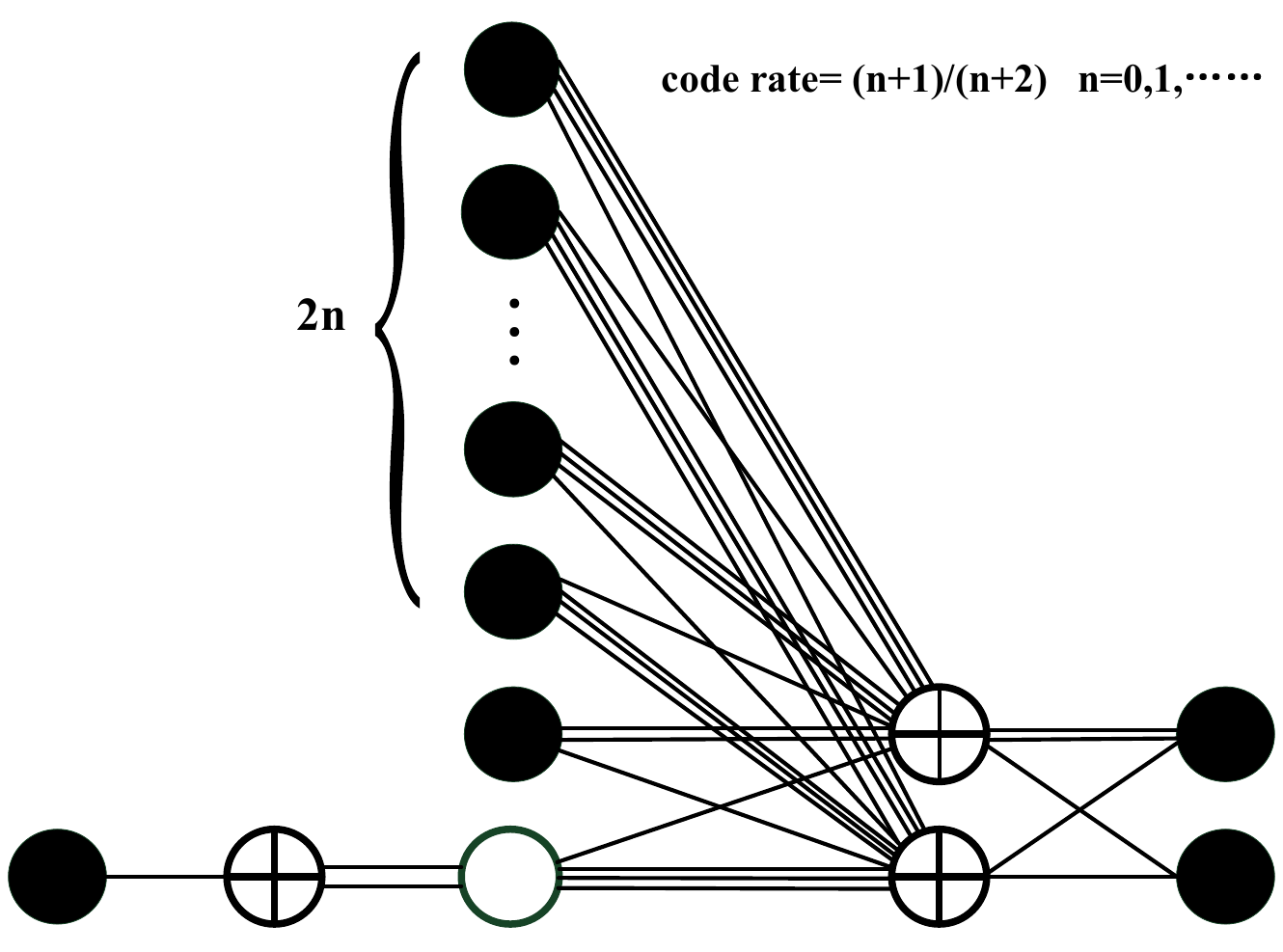}
  \caption{The AR4JA family of protographs with rates 1/2 and higher.}
  \label{fig:protograph_AR4JA}
\end{figure}

Low-density parity-check (LDPC) codes are  well-known error correction codes due to the
capacity-approaching and parallel decoding properties. In this paper, we consider a sub-type of LDPC codes known as the protograph based LDPC codes, which have excellent error performance and low complexity. Fig. \ref{fig:protograph_AR4JA} shows the AR4JA family of protographs with rates 1/2 and higher \cite{Divsalar2009Capacity}, where the black circles and the white circles with cross represent the variable nodes (VNs) and the check nodes (CNs), respectively; and the white circles indicate the punctured VNs. The protograph can also be described by the base matrix $\mathbf{B}=(b_{i,j})$, where the value of the entry $b_{i,j}$ indicates the number of edges connecting the \textit{i}th CN and the \textit{j}th VN. Eq. (\ref{eq:1/2 matrix}) and Eq. (\ref{eq:4/5 matrix}) show the base matrices of rate 1/2 and 4/5 AR4JA codes, corresponding to the protographs with $n=0$ and $n=3$ in Fig. \ref{fig:protograph_AR4JA}. These two types of AR4JA codes are adopted for error correction over DNA channels with the Nanopore sequencer and the Illumina sequencer, respectively. The parity-check matrix is generated by lifting the base matrix, which can be summarized by “copy-and-permute” \cite{Thorpe2003Low}. For the sake of simplicity, in this work, we consider conventional belief propagation (BP) algorithm \cite{Chen2002BP} for the decoding of LDPC codes.

\begin{equation}    
{\mathbf{B}_{AR4JA\_1/2}}\!=\!{\left[           
   \begin{array}{llll}   
    1\ 2\ 0\ 0\ 0\\
    0\ 3\ 1\ 1\ 1\\
    0\ 1\ 2\ 2\ 1\\
  \end{array}
  \label{eq:1/2 matrix}
   \right ]} 
\end{equation}

\begin{equation}    
{\mathbf{B}_{AR4JA\_4/5}}\!=\!{\left[           
   \begin{array}{lllllllll}   
    1\ 2\ 0\ 0\ 0\ 0\ 0\ 0\ 0\ 0\ 0\\
    0\ 3\ 1\ 3\ 1\ 3\ 1\ 3\ 1\ 1\ 1\\
    0\ 1\ 2\ 1\ 3\ 1\ 3\ 1\ 3\ 2\ 1\\
  \end{array}
  \label{eq:4/5 matrix}
   \right ]} 
\end{equation}

\section{Two Asymmetric DNA Sequencing Channels} 
\label{sec:Sequencing Channels}
In DNA based data storage, errors might occur at any stage, including DNA synthesis, sample preparing, storage, and DNA sequencing. In this work, we focus on the errors arising in the DNA sequencing process. In the following, we discuss two error models of two widely utilized sequencing techniques. 

\begin{figure}[htbp]
  \centering
  \includegraphics[ width=\figwidth]{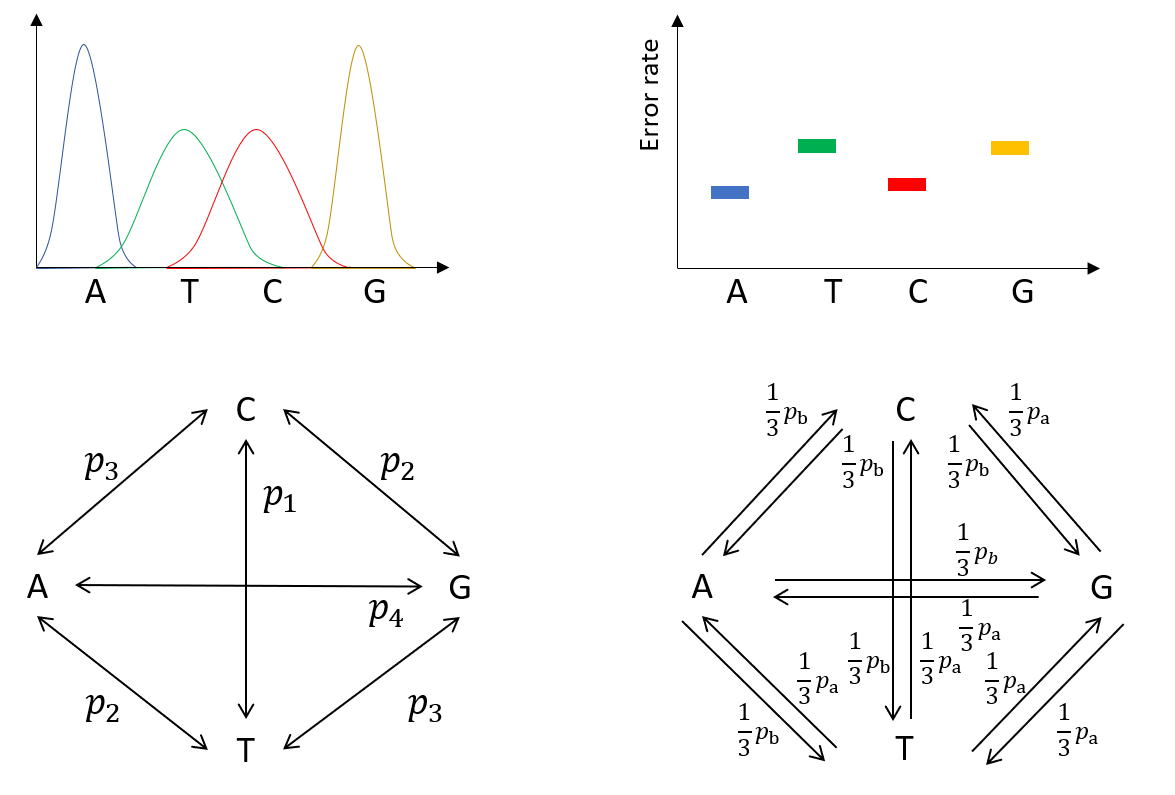}
  \caption{Two asymmetric sequencing channel models. The left two refer to the Nanopore sequencing channel while the right two refer to the Illumina sequencing channel.}
  \label{fig:channel}
\end{figure}

In the Nanopore sequencing process, the DNA strands migrate through the Nanopore at a constant rate, and only one nucleotide of the DNA strands is read  at a given time. Due to the different atomic structures, each nucleotide can be detected by observing the changes of the ionic current drop while the DNA strand passing through the pore. The current drop response of the Nanopore sequencer was reported in \cite{gabrys2017asymmetric} (the top left of Fig. \ref{fig:channel}). As can be seen, the mutation probability between the nucleotides 'T' and 'C' is the most significant; and the nucleotides 'A' and 'G' seem much less likely to mutate from each other. Based on these observations,  we model the Nanopore sequencing as an asymmetric substitution error model (the bottom left of Fig. \ref{fig:channel}). Note that the error model is similar to \cite{Zhiying2019}, while consists of four different error possibilities represented by $p_1, p_2, p_3, p_4$, defined as $p_1=4\alpha, p_2=\alpha, p_3=0.01, p_4\approx0$, where $\alpha\in(0,\frac{1}{4})$, and $p_4\ll p_3<p_2<p_1$.


From the experimental error analysis of Illumina's NextSeq in \cite{Organick2018}, we find that asymmetric errors also exist, in which the substitution error probability of 'T' and 'G' ($p_a$) is higher than the substitution error probability of 'A' and 'C' ($p_b$). Accordingly, we build the asymmetric substitution error model for the Illumina sequencer with the assumption of equal mutation probabilities from one nucleotide to the other three nucleotides as shown in the bottom right of Fig. \ref{fig:channel}. The substitution possibilities $p_a, p_b$ follow $p_a=1.5\beta, p_b=\beta$, where $\beta$ satisfies $\beta\in[0,\frac{2}{3}]$.

\section{Proposed Hybrid Coding Architecture} 
\label{sec:Heterogeneous Coding Scheme}

\begin{figure*}[htbp]
  \centering
  \includegraphics[width=15.2 cm, height = 6 cm]{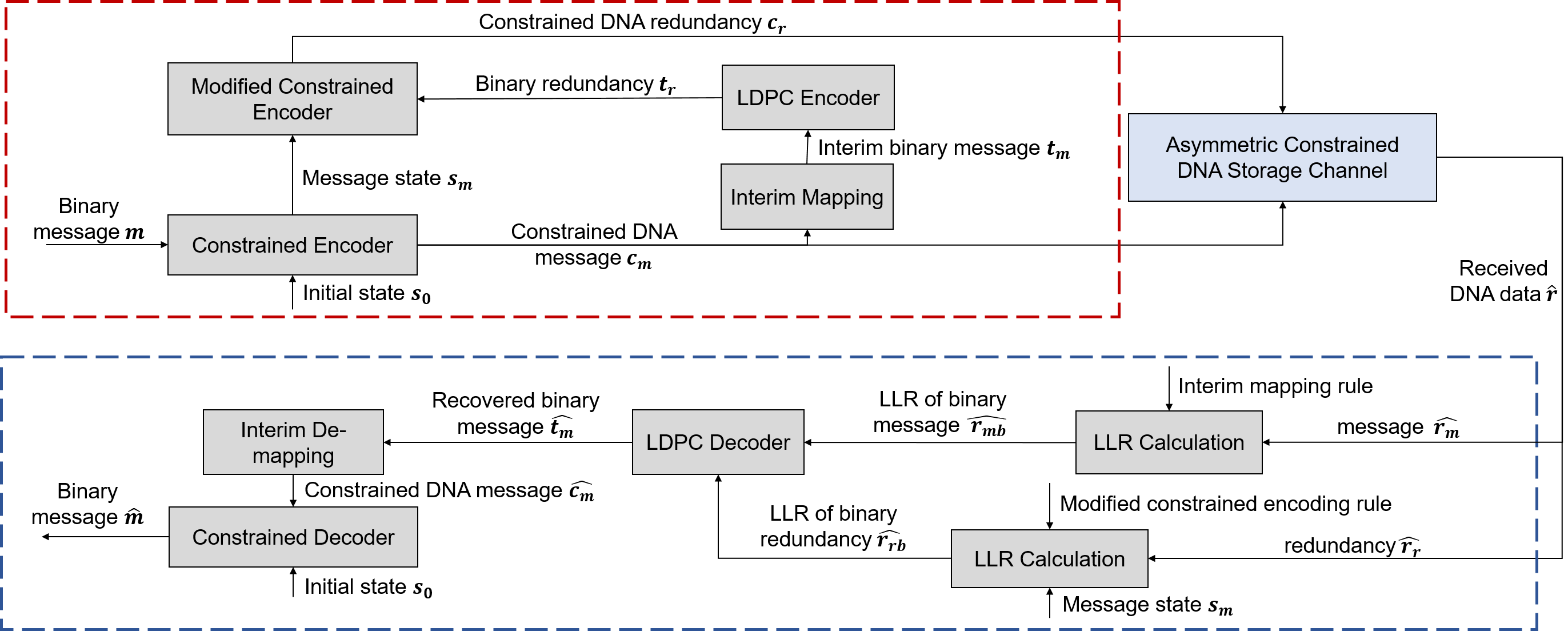}
  \caption{Proposed hybrid encoding/decoding for DNA data storage.}
  \label{fig:system}
\end{figure*}

In this section, we describe the proposed hybrid coding architecture as shown in Fig. \ref{fig:system}. The proposed scheme writes the source data into the homopolymer-constrained DNA record (red block), while protecting the encoded data from the asymmetric substitution errors occurring in the process of DNA sequencing upon a retrieval request (blue block). 

\subsection{Encoding}
We input the source binary message $\bm{m}$ into the constrained encoder, i.e., VL-RLL encoder based on Table. \ref{tab:VL-RLL} and obtain an output DNA message sequence $\bm{c_m}$ consisting of four nucleotide alphabets, i.e., A/T/C/G. The output sequence $\bm{c_m}$ is reserved before sending to the storage. Meanwhile, the sequence $\bm{c_m}$ is mapped into a binary message sequence $\bm{t_m}$ via an interim mapping that directly maps 'A' to '00', 'T' to '01', 'G' to '10', and 'C' to '11'. The mapping is chosen to keep consistent with the VL-RLL encoding in which the last step converts the quaternary symbols to DNA symbols according to 0$\to$A, 1$\to$T, 2$\to$G, and 3$\to$C. After the interim mapping, a systematic binary LDPC encoding is performed based on the mapped binary message $\bm{t_m}$ to generate  redundant  binary sequence $\bm{t_r}$. Next, we convert the redundant sequence into a form that complies with the homopolymer (run-length) constraint for  further storage.  Through a modified constrained encoder, i.e., modified VL-RLL encoder based on Table. \ref{tab:M-VL-RLL}, the binary redundancy sequence $\bm{t_r}$ is converted into constrained DNA redundancy sequence $\bm{c_r}$, and then attaching at the right of the reserved constrained DNA message $\bm{c_m}$. The rationale behind choosing Table. \ref{tab:M-VL-RLL} is discussed in Remarks. Note that a message state $\bm{s_m}$, that indicates the last alphabet of the information block $\bm{c_m}$, is fed to the encoding process of the binary redundancy sequence $\bm{t_r}$, to ensure that the resultant DNA sequence consisting of $\bm{c_r}$ and $\bm{c_m}$ satisfy the homopolymer constraint. The constructed DNA sequences are then synthesized and stored in the DNA data storage. 

%


Instead of performing LDPC encoding before constrained mapping (which is commonly used in existing work), we move the LDPC encoding after the constrained mapping of the source bits. This step enables us to correct the channel errors that occur on the constrained codewords (which are the exact stored entity) before the errors diffuse to the recovered source data in the reverse of the precoding and mapping step that sets basic of most of the constrained codes. This design pipeline exhibits much importance as the error-propagation is much severer in variable-length constrained codes (as we have used) than the conventional block codes. For the sake of simplicity, we consider binary LDPC code instead of quaternary LDPC code and hence use an interim mapping (to convert quaternary data to binary data) on the constrained DNA data. In other words, with quaternary LDPC code, we need to transform the unrestricted quaternary redundancy to the constraint-satisfying quaternary redundancy, which might incur more computation cost than the binary case.

\textbf{Remarks:} In the modified VL-RLL mapping, we only change the last mapping of 11111$\to$003 in the original VL-RLL mapping (Table. \ref{tab:VL-RLL}) to 11111X$\to$003 (Table. \ref{tab:M-VL-RLL}), where 'X' represents either '0' or '1'. By using this fuzzy bit in the source word while mapping to a fixed transition word, we avoid the synchronization problem that may arise in the original VL-RLL mapping (i.e., all bijections have a rate 2 bits/symbol except the last one with rate $5/3$ bits/symbol) at the cost of sacrificing the accuracy of de-mapping this fuzzy bit. Note that, we only use this mapping for the redundancy bits generated by the LDPC codes. 


\begin{table}
\setlength{\extrarowheight}{1.5mm} 
\setlength{\tabcolsep}{0.8mm}
\centering 
\caption{Modified VL-RLL Mapping Rule}\label{tab:M-VL-RLL}
\begin{threeparttable}
\begin{tabular}{|c|c|c|c|c|c|c|c|c|c|}
\hline
\textbf{Source word} &  $00$ & $01$ &  $10$ & $1100$ &  $1101$ & $1110$ &  $111100$ & $111101$ & $11111X$\\ 
\hline
\textbf{Transition word} &  $1$ & $2$ &  $3$ & $01$ &  $02$ & $03$ &  $001$ & $002$ & $003$\\
\hline
\end{tabular}
\end{threeparttable}
\end{table}

\subsection{Decoding}
To recover the source data from the storage, we perform decoding on the received DNA data $\bm{\hat{r}}$ from the DNA sequencing process. With the assumption of the knowledge of the boundary between information blocks and redundancy blocks in the decoder, the received DNA data are first separated into message block $\bm{\widehat{r_m}}$, denoted by $r_{m_1}r_{m_2}...r_{m_{i-1}}r_{m_i}$ and redundancy block $\bm{\widehat{r_r}}$, denoted by $r_{r_1}r_{r_2}...r_{r_{j-1}}r_{r_j}$. This is because the nucleotide alphabets in the two blocks, i.e., $r_{m_i}$ and $r_{r_j}$, pass the channel information to the encoded bits in the LDPC codeword in different ways. 

Specifically, for \textit{i}th received nucleotide $r_{m_i}$ in the message block $\bm{\widehat{r_m}}$, its associative received binary bits $(b^1_i, b^2_i)$ in the LDPC codeword is determined by the interim mapping. The received alphabet $r_{m_i}$ with the channel information, supplies the initial LLRs $(\mathcal{L}^0_{i^1}, \mathcal{L}^0_{i^2})$ to the correlative two bits, facilitating the LDPC decoding using BP algorithm. However, for \textit{j}th received nucleotide $r_{r_j}$ in the redundancy block $\bm{\widehat{r_r}}$, finding the associative received binary bits are not as straightforward as in the information block as the nucleotide alphabets are encoded via a constrained encoding, in which the precoding process implicates that two neighboring nucleotide alphabets in the constrained codeword jointly determine a relevant transition symbol before de-mapping to the associative binary bits. 

Based on Table. \ref{tab:M-VL-RLL}, it is found that except symbol '3', each transition symbol in the transition words is uniquely mapped to two binary bits in the source word (i.e., 1$\to$00, 2$\to$01, 0$\to$11). The transition symbol '3' is involved in three transition words, i.e., '3', '03', '003'. In transition words '3' and '03', the symbol '3' is mapped to the binary bits '10' in the corresponding source words. However, in the case of '003', '3' can be mapped to either '10' or '11' in the source word '11111X'. With the i.i.d assumption of binary bits, we can derive the probabilities $p_w$ for '3' mapping to '10',
\begin{equation} \label{eq:weight}
\begin{aligned}
p_w=\frac{p_{10}+p_{1110}+\frac{1}{2}p_{11111X}}{p_{10}+p_{1110}+p_{11111X}}=\frac{2^{-2}+2^{-4}+2^{-7}}{2^{-2}+2^{-4}+2^{-6}}=\frac{41}{42}
\end{aligned}
\end{equation}then $1-p_w = \frac{1}{42}$ for '3' mapping to '11'.

In the following, we introduce how the asymmetric channel information is passed to the initial LLR of each encoded binary bit in the LDPC codewords. We first explain the LLRs of binary bits that relate to the received information block $\bm{\widehat{r_m}}$. For a pair of \textit{i}th received nucleotide alphabet $r_{m_i}$ in $\bm{\widehat{r_m}}$ and \textit{i}th stored nucleotide alphabet $c_{m_i}$ in $\bm{c_m}$, we have $\Pr(x_i=c_{m_i}|y_i=r_{m_i})$ indicating the event possibility, which is represented by the substitution possibility in the channel models in Fig. \ref{fig:channel}. The initial LLRs $(\mathcal{L}^0_{i^1}, \mathcal{L}^0_{i^2})$ of the associative $(b^1_i, b^2_i)$ thus can be estimated on the basis of the interim mapping, i.e., A$\to$00, T$\to$01, G$\to$10, C$\to$11. The initial LLR of the relevant bit is derived by,
\begin{equation*} 
\mathcal{L}^0_{i^k}=\log\frac{\Pr(b^k_i=0|y_i=r_{m_i})}{\Pr(b^k_i=1|y_i=r_{m_i})}
\end{equation*}where $k \in \{1, 2\}$. An example for the Nanopore channel is shown as below.
If $r_{m_i}=A$, we have,
\begin{equation*}   
\begin{aligned}
\mathcal{L}^0_{i^1}&=\log\frac{\Pr(x_i=A|y_i=A)+\Pr(x_i=T|y_i=A)}{\Pr(x_i=G|y_i=A)+\Pr(x_i=C|y_i=A)}\\
&=\log\frac{(1-p_2-p_3-p_4)+p_2}{p_4+p_1}\\
\mathcal{L}^0_{i^2}&=\log\frac{\Pr(x_i=A|y_i=A)+\Pr(x_i=G|y_i=A)}{\Pr(x_i=T|y_i=A)+\Pr(x_i=C|y_i=A)}\\
&=\log\frac{(1-p_2-p_3-p_4)+p_4}{p_2+p_3}
\end{aligned}
\end{equation*}The LLRs of received 'T', 'C' and 'G' in the information blocks can be estimated in the similar way.

\comment{
Similarly, if $r_{m_i}=T$, we have, 
\begin{equation*}
\begin{aligned}
\mathcal{L}^0_{i^1}&=\log\frac{\Pr(x_i=A|y_i=T)+\Pr(x_i=T|y_i=T)}{\Pr(x_i=G|y_i=T)+p(x_i=C|y_i=T)}\\
&=\log\frac{p_2+(1-p_1-p_2-p_3)}{p_3+p_1}\\
\mathcal{L}^0_{i^2}&=\log\frac{\Pr(x_i=A|y_i=T)+\Pr(x_i=G|y_i=T)}{\Pr(x_i=T|y_i=T)+\Pr(x_i=C|y_i=T)}\\
&=\log\frac{p_2+p_3}{(1-p_1-p_2-p_3)+p_1}
\end{aligned}
\end{equation*}
}

Next, we explain the LLRs associated with the redundancy block $\bm{c_r}$. As $\bm{c_r}$ is precoded from the transition sequence that consists of transition words via a differential operation, one corrupted nucleotide in the received redundancy block $\bm{\widehat{r_r}}$ induces two erroneous transition symbols in the transition sequence, potentially resulting in burst errors in the corresponding binary redundancy $\bm{t_r}$ after performing the reverse of the mapping rule in Table. \ref{tab:M-VL-RLL}. In other words, each transition symbol that relates to two bits in the parity-check of the LDPC codeword is relevant to two neighboring nucleotide alphabets in the received constrained redundancy block $\bm{\widehat{r_r}}$. Thus, we consider two received nucleotides one time for determining the transition symbol before passing the initial LLRs to two parity-check bits in the LDPC codeword based on Table. \ref{tab:M-VL-RLL}. The event possibility thus relies on the substitution possibilities of each pair of neighboring nucleotides (\textit{(j-1)}th and \textit{j}th) in $\bm{t_r}$, representing by $\Pr(x_{j-1}x_j=c_{r_{j-1}}c_{r_{j}}|y_{j-1}y_j=r_{r_{j-1}}r_{r_j})$. The initial LLRs $(\mathcal{L}^0_{j^1}, \mathcal{L}^0_{j^2})$ thus can be estimated on the basis of the modified VL-RLL mapping as shown in Table.~\ref{tab:M-VL-RLL}. The initial LLR of the relevant bit is computed by,
\begin{equation*} 
\mathcal{L}^0_{j^k}=\log\frac{\Pr(b^k_j=0|y_{j-1}y_j=r_{r_{j-1}}r_{r_{j}})}{\Pr(b^k_j=1|y_{j-1}y_j=r_{r_{j-1}}r_{r_{j}})}
\end{equation*}where $k \in \{1, 2\}$. In below, we show an example.
If we receive $r_{r_{j-1}}r_{r_{j}}=TC$, then
\begin{equation*}   
P1=\Pr(b^1_j=0|y_{j-1}y_j=TC)=\Pr(x_{j-1}x_j|y_{j-1}y_j=TC)
\end{equation*} where $x_{j-1}x_j=CT, CA, AG, AT, TC, TG, GA, GC$, all neighboring pairs produce transition symbols either '1' or '2'. Meanwhile, 
\begin{equation*}  
P2=\Pr(b^1_j=1|y_{j-1}y_j=TC)=\Pr(x_{j-1}x_j|y_{j-1}y_j=TC)
\end{equation*} where $x_{j-1}x_j=CC, CG, AA, AC, TT, TA, GG, GT$, all neighboring pairs produce transition symbols either '0' or '3'. Similarly, we have,
\begin{equation*}   
P3=\Pr(b^2_j=0|y_{j-1}y_j=TC)=\Pr(x_{j-1}x_j|y_{j-1}y_j=TC)
\end{equation*} where $x_{j-1}x_j=CG, CA, AC, AT, TA, TG, GT, GC$, all neighboring pairs produce transition symbols either '1' or '3'. And
\begin{equation*}  
P4=\Pr(b^2_j=1|y_{j-1}y_j=TC)=\Pr(x_{j-1}x_j|y_{j-1}y_j=TC)
\end{equation*} where $x_{j-1}x_j=CC, CT, AA, AG, TT, TC, GG, GA, CG, \\AC, TA, GT$, all neighboring pairs produce transition symbols from $\{$'0', '2', '3'$\}$. Therefore, we have,
\begin{equation*}
\begin{split}
P1=&\Pr(C|T)\cdot(\Pr(T|C)+\Pr(A|C))+\Pr(A|T)\cdot\\
&(\Pr(G|C)+\Pr(T|C))+\Pr(T|T)\cdot(\Pr(C|C)\\
&+\Pr(G|C))+\Pr(G|T)\cdot(\Pr(A|C)+\Pr(C|C))
\end{split}
\end{equation*}
\begin{equation*}
\begin{split}
P2=&\Pr(C|T)\cdot(\Pr(C|C)+\Pr(G|C))+\Pr(A|T)\cdot\\
&(\Pr(A|C)+\Pr(C|C))+\Pr(T|T)\cdot(\Pr(T|C)\\
&+\Pr(A|C))+\Pr(G|T)\cdot(\Pr(G|C)+\Pr(T|C))
\end{split}
\end{equation*}
\begin{equation*}
\begin{split}
P3=&\Pr(C|T)\cdot(p_w\cdot\Pr(G|C)+\Pr(A|C))+\Pr(A|T)\\
&\cdot(p_w\cdot\Pr(C|C)+\Pr(T|C))+\Pr(T|T)\cdot(p_w\cdot\\
&\Pr(A|C)+\Pr(G|C))+\Pr(G|T)\cdot(p_w\cdot\Pr(T|C)\\
&+\Pr(C|C))
\end{split}
\end{equation*}
\begin{equation*}
\begin{split}
P4=&\Pr(C|T)\cdot(\Pr(C|C)+\Pr(T|C)+(1-p_w)\\
&\cdot\Pr(G|C))+\Pr(A|T)\cdot(\Pr(A|C)+\Pr(G|C)\\
&+(1-p_w)\cdot\Pr(C|C))+\Pr(T|T)\cdot(\Pr(T|C)\\
&+\Pr(C|C)+(1-p_w)\cdot\Pr(A|C))+\Pr(G|T)\\
&\cdot(\Pr(G|C)+\Pr(A|C)+(1-p_w)\cdot\Pr(T|C))
\end{split}
\end{equation*}Note that $p_w$ is the possibility of transition symbol '3' mapped to binary bits '10', calculating by Eq. (\ref{eq:weight}). As a result, we obtain,
\begin{equation*}   
\mathcal{L}^0_{j^1}=\log\frac{P1}{P2};\ \mathcal{L}^0_{j^2}=\log\frac{P3}{P4}
\end{equation*}

\comment{
\begin{equation*}   
\begin{split}
\mathcal{L}^0_{j^1}=\log\frac{P1}{P2}=\log\frac{p_1\cdot(p_1+p_3)+p_2\cdot(p_2+p_1)+(1-p_1-p_2-p_3)\cdot((1-p_1-p_2-p_3)+p_2)+p_3\cdot(p_3+(1-p_1-p_2-p_3))}{p_1\cdot((1-p_1-p_2-p_3)+p_2)+p_2\cdot(p_3+(1-p_1-p_2-p_3))+(1-p_1-p_2-p_3)\cdot(p_1+p_3)+p_3\cdot(p_2+p_1)}\\
\mathcal{L}^0_{j^2}=\log\frac{P3}{P4}=\log\frac{}{}
\end{split}
\end{equation*}
}

With the above LLRs, we perform the standard belief propagation (BP) decoding. 

\section{Numerical Results and Discussion} 
\label{sec:Results}

We fix the original information block lengths as 300 bits and 1000 bits for the Illumina and Nanopore channel models, respectively. Considering the different error probabilities of the Nanopore sequencing ($\sim10^{-2}$) and the Illumina sequencing ($\sim10^{-3}$), the coding rates of $\mathcal{R}_{nano}=1/2$ and $\mathcal{R}_{illu}=4/5$ are adopted correspondingly. However, in Illunima case, after the VL-RLL and interim mapping, the encoding data change with continuous even lengths from 300 bits to 318 bits, which can not all be matched by the fixed size of $\mathbf{B}_{AR4JA\_{4/5}}$ ($3\times11$). Thus, a few rows and columns of the matched $\mathbf{H}$ matrices are punctured to generate some approximately $4/5$ rate $\mathbf{H}$ matrices for certain block lengths. For instance, for encoding 302 data bits, the $\mathbf{H}$ matrix with a rate of approximate $4/5$ can be obtained by puncturing the last three columns and last one row of the $\mathbf{H}$ matrix with size of $114 \times 418$ after 38 times of lifting from $ \mathbf{B}_{AR4JA\_{4/5}}$ (which is corresponding to the encoding of 304 data bits). Note that we use practical channel parameters for the simulations, i.e., $\alpha \in[0.03, 0.04]$ \cite{Zhiying2019} and $\beta \in[0.5\times10^{-3}, 1.5\times10^{-3}]$ \cite{Organick2018}.

\begin{figure}[htbp]
  \centering
  \includegraphics[width=\figwidth]{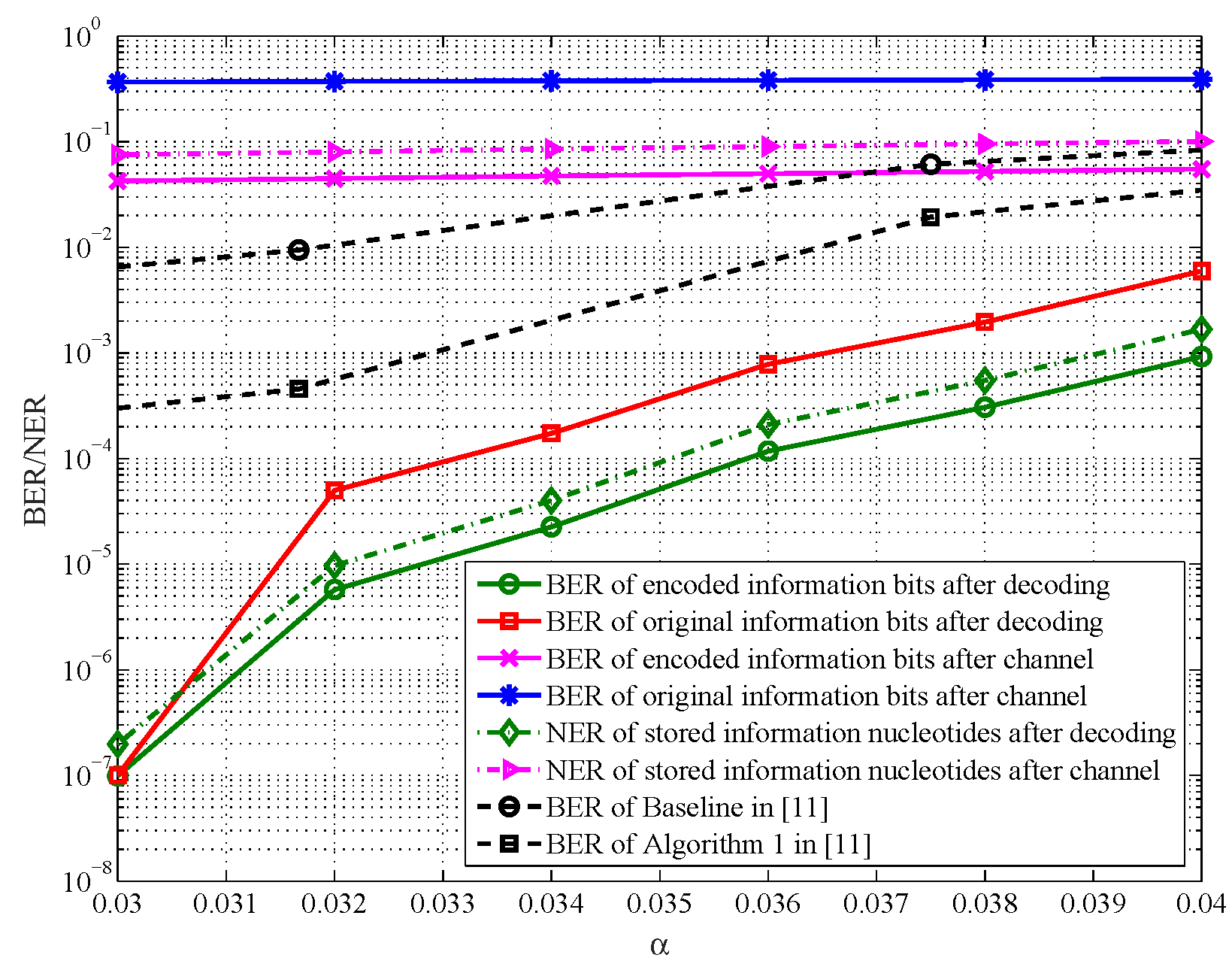}
  \caption{The BER/NER performance of rate $\frac{1}{2}$ AR4JA codes over the Nanopore sequencing channel}
  \label{fig:nanopore_BER}
\end{figure}

\begin{figure}[htbp]
  \centering
  \includegraphics[width=\figwidth]{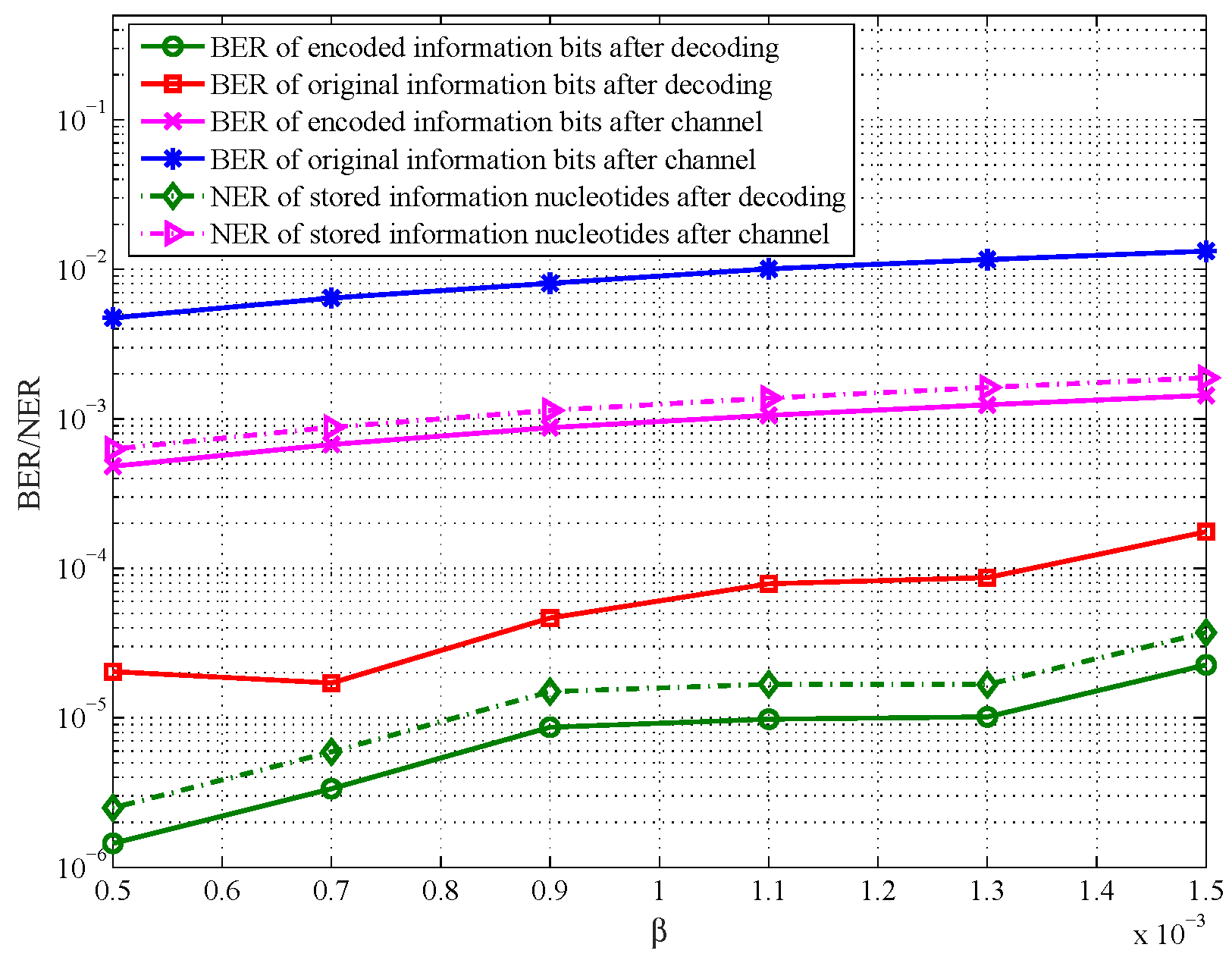}
  \caption{The BER/NER performance of rate $\frac{4}{5}$ AR4JA codes over the Illumina sequencing channel}
  \label{fig:illumina_BER}
\end{figure}

The BER (bit error rate) and NER (nucleotide error rate) performance of two channels are shown in Fig. \ref{fig:nanopore_BER} and Fig. \ref{fig:illumina_BER}. Comparing the solid blue lines with the solid magenta lines, we can observe an increase of error rate. This is due to the constrained decoding process, i.e., the reverse of VL-RLL. As discussed, one nucleotide in error might lead to two transition symbols in error when the reverse of precoding is performed. Furthermore, an erroneous transition symbol might lead to severe error propagation in the subsequent bits in the reverse of mapping based on Table. \ref{tab:VL-RLL}. For instance, when a transition word '002' is mutated to '003' due to the erroneous nucleotide in the channel, the original 6-bit '111101' is wrongly de-mapped to 5-bit '11111', leading to all subsequent mapped bits having synchronization problems against the original source bits. 



The solid green lines represent the BER of the encoded information bits ($\bm{\widehat{t_m}}$ against $\bm{t_m}$ in Fig. \ref{fig:system}) after LDPC decoding. The solid red lines represent the BER of the original information bits ($\bm{\hat{m}}$ against $\bm{m}$ in Fig.\ref{fig:system}) after LDPC decoding and constrained decoding. As shown in both figures, less error is observed than the case before LDPC decoding. Notice that when the BER of the green line is approaching 0, the BER of the red line is also approaching 0, which indicates that errors are corrected by LDPC decoding at the near-nucleotide-level, reducing the error propagation in the process of constrained decoding. The dotted magenta lines show the NER of the information block in the sequencing channels with different parameters ($\bm{c_m}$ against $\bm{\widehat{r_m}}$ in Fig. \ref{fig:system}). The dotted green lines show the NER of the information block after LDPC decoding while before constrained decoding ($\bm{c_m}$ against $\bm{\widehat{c_m}}$ in Fig. \ref{fig:system}). As expected, the trends of the two dotted lines are consistent with their relevant BER lines. 

\textbf{Comparison with \cite{Zhiying2019}:} Different from \cite{Zhiying2019} which focuses on devising intelligent decoding algorithms, our work focuses on practical code design for DNA data storage. In \cite{Zhiying2019}, the authors simply mapped 2-source bits to 1 nucleotide symbol and hence the resultant DNA sequences might not satisfy the biochemical constraint. In contrast, we propose a more practical coding scheme with efficient decoding for error resilience in DNA data storage, where the resultant DNA sequences satisfy the biochemical constraint, potentially benefiting the storage process (i.e., less errors in DNA sequencing). In addition, we compare the error rate performance of LDPC codes with \cite{Zhiying2019} under the same average error rate in the Nanopore sequencing channel, where we set $3.5\lambda$ of \cite{Zhiying2019} equal to $3\alpha+0.01$ of this work. The dotted black lines in Fig.~\ref{fig:nanopore_BER} represent two BER (baseline BER and Algorithm 1 BER) performances of \cite{Zhiying2019}. The baseline scheme uses the conventional BP decoding (similar to this work), while Algorithm 1 uses the BP decoding with side information. We observe that our scheme outperforms both BER performances even by using the conventional BP decoding. Note that a better BER performance can be achieved with our scheme by utilizing side information at the decoder (i.e., similar to the algorithms presented in \cite{Zhiying2019}).






\textbf{Coding potential:} We analyze the coding potential by using VL-RLL and modified VL-RLL for constrained encoding the information bits and the parity-check bits of the LDPC codewords, respectively. As discussed, VL-RLL offers an average coding potential of $\mathbf{R_{info}}=1.976$ bits/nt based on Eq. (\ref{eq:code rate}). Meanwhile, based on Table. \ref{tab:M-VL-RLL}, the coding potential of the modified VL-RLL becomes $\mathbf{R_{red}}\simeq 2$ bits/nt leveraging by the fuzzy bit. Thus, the average overall coding potential becomes
\begin{equation*}   
\mathbf{R_{all}}=\frac{m+\frac{2m(1-\mathcal{R})}{\mathbf{R_{info}}\cdot\mathcal{R}}}{\frac{m}{\mathbf{R_{info}}}+\frac{2m\cdot(1-\mathcal{R})}{\mathbf{R_{info}}\cdot\mathbf{R_{red}}\cdot\mathcal{R}}}
\end{equation*}where $m$ is the length of the original information bits, $\mathcal{R}$ is the code rate of LDPC.
Thus, we obtain the coding potential $\mathbf{R_{all}}\simeq (2-0.024\mathcal{R}$). For Nanopore case where $\mathcal{R}_{nano}=1/2$, the coding potential becomes $\sim1.988$ bits/nt; and for Illumina case where $\mathcal{R}_{illu}=4/5$, the coding potential becomes $\sim1.981$ bits/nt, presenting only 1\% gap from the upper boundary $2$ bits/nt. The achieved coding potential is higher than the reported in the existing works \cite{Immink2018, song2018codes, 8695057}.

%
%

\section{Conclusion}\label{sec:Conclusion}
We have introduced a hybrid coding architecture, which can correct asymmetric errors in the DNA sequencing processes while satisfying the biochemical constraint desired for the processes, offering a practical code design for DNA data storage. The VL-RLL code is used with lower complexity and higher code potential than the conventional block constrained codes. The LDPC code is incorporated at a near-nucleotide level to potentially hamper the occurrence of the error-propagation in the reverse of VL-RLL encoding. Moreover, a modified VL-RLL code is developed for the constrained mapping of the redundant bits generated by the LDPC encoder, which facilitates an effective decoding process and a very high coding potential ($\sim2$). The simulation results show that the proposed scheme can tackle the asymmetric substitution errors caused by the Nanopore and Illumina sequencers. In the future, we aim to work on the design of optimized LDPC codes for the asymmetric channels discussed in the paper.

\small
\bibliography{reference_new}
\bibliographystyle{IEEEtran}

\end{document}